\documentclass{aa}  
\usepackage{color}
\usepackage{graphicx}
\usepackage{txfonts}

\newcommand{\ms}{m\,s$^{-1}$}

\def\vsini{\ensuremath{{\upsilon}\sin i}}
\def\kms{$\mathrm{km\,s}^{-1}$}
\def\ms{\hbox{\,m\,s$^{-1}$}}         %m.s -1
\def\m2s2{\hbox{\,m$^{2}$\,s$^{-2}$}} %m2.s -2
\def\kms{\hbox{\,km\,s$^{-1}$}}       %km.s -1
\def\vsini{\hbox{$v$\,sin\,$i$}}      %vsini
\begin{document} 
   \title{The GAPS Programme at TNG}
   \subtitle{XXXIII. HARPS-N detects multiple atomic species in emission from the dayside of KELT-20b\thanks{Based on observations made with the Italian Telescopio Nazionale Galileo (TNG) operated on the island of La Palma by the Fundacion Galileo Galilei of the INAF at the Spanish Observatorio Roque de los Muchachos of the IAC in the frame of the program Global Architecture of the Planetary Systems (GAPS).}
}
   \titlerunning{Emission from the day-side of KELT-20b}
   \authorrunning{F.~Borsa et al.}
 %  \subtitle{I. Overviewing the $\kappa$-mechanism}
    \author{F.~Borsa\inst{\ref{brera}}, 
      P.~Giacobbe\inst{\ref{torino}}, 
      A.~S.~Bonomo\inst{\ref{torino}},  
      M.~Brogi\inst{\ref{warwick},\ref{torino},\ref{warwick2}}, 
      L.~Pino\inst{\ref{arcetri}},
      L.~Fossati\inst{\ref{graz}}, 
      A.~F.~Lanza\inst{\ref{catania}},
      V.~Nascimbeni\inst{\ref{padova}},  
      A.~Sozzetti\inst{\ref{torino}}, 
      F.~Amadori\inst{\ref{torino}}, 
      S.~Benatti\inst{\ref{palermo}}, 
      K.~Biazzo\inst{\ref{monteporzio}}, 
      A.~Bignamini\inst{\ref{trieste}}, 
      W.~Boschin\inst{\ref{tng}, \ref{IAC},\ref{IAC2}},
      R.~Claudi\inst{\ref{padova}},  
       R.~Cosentino\inst{\ref{tng}}, 
       E.~Covino\inst{\ref{napoli}}, 
       S.~Desidera\inst{\ref{padova}}, 
       A.~F.~M.~Fiorenzano\inst{\ref{tng}}, 
      G.~Guilluy\inst{\ref{torino}}, 
       A.~Harutyunyan\inst{\ref{tng}}, 
       A.~Maggio\inst{\ref{palermo}},    
      J.~Maldonado\inst{\ref{palermo}}, 
      L.~Mancini\inst{\ref{romauniv},\ref{heidelberg},\ref{torino}},  
       G.~Micela\inst{\ref{palermo}},  
       E.~Molinari\inst{\ref{cagliari}}, 
       M.~Molinaro\inst{\ref{trieste}}, 
       I.~Pagano\inst{\ref{catania}}, 
       M.~Pedani\inst{\ref{tng}}, 
       G.~Piotto\inst{\ref{padovauniv}}, 
       E.~Poretti\inst{\ref{brera},\ref{tng}},    
       M.~Rainer\inst{\ref{arcetri}},  
       G.~Scandariato\inst{\ref{catania}},  
       H.~Stoev\inst{\ref{tng}}
           }
   \institute{INAF -- Osservatorio Astronomico di Brera, Via E. Bianchi 46, 23807 Merate (LC), Italy \label{brera}
   \and
INAF -- Osservatorio Astrofisico di Torino, Via Osservatorio 20, 10025, Pino Torinese, Italy  \label{torino}
\and
Department of Physics, University of Warwick, Coventry CV4 7AL, UK  \label{warwick}
\and
Centre for Exoplanets and Habitability, University of Warwick, Gibbet Hill Road, Coventry CV4 7AL, UK  \label{warwick2}
   \and
   INAF -- Osservatorio Astrofisico di Arcetri, Largo E. Fermi 5, 50125 Firenze, Italy \label{arcetri}
\and
Space Research Institute, Austrian Academy of Sciences, Schmiedlstrasse 6, A-8042 Graz, Austria  \label{graz}
\and
   INAF -- Osservatorio Astrofisico di Catania, Via S.Sofia 78, 95123, Catania, Italy  \label{catania}
   \and
INAF -- Osservatorio Astronomico di Padova, Vicolo dell'Osservatorio 5, 35122, Padova, Italy  \label{padova}
\and
INAF -- Osservatorio Astronomico di Palermo, Piazza del Parlamento, 1, 90134, Palermo, Italy  \label{palermo}
\and
INAF – Osservatorio Astronomico di Roma, via Frascati 33, I00040, Monte Porzio Catone (RM), Italy\label{monteporzio}
\and
INAF -- Osservatorio Astronomico di Trieste, via Tiepolo 11, 34143 Trieste, Italy  \label{trieste}
\and
Fundaci{\'o}n Galileo Galilei - INAF, Rambla Jos{\'e} Ana Fernandez P{\'e}rez 7, 38712 Bre$\tilde{\rm n}$a Baja, TF - Spain  \label{tng}
\and
Instituto de Astrof\'{\i}sica de Canarias (IAC), C/V\'{\i}a L\'actea s/n, 38205 La Laguna, TF - Spain \label{IAC} 
\and 
Departamento de Astrof\'{\i}sica, Universidad de La Laguna (ULL), 38206 La Laguna, TF - Spain \label{IAC2} 
\and
INAF -- Osservatorio Astronomico di Capodimonte, Salita Moiariello 16, 80131, Napoli, Italy  \label{napoli}
\and 
Department of Physics, University of Rome Tor Vergata, Via della Ricerca Scientifica 1, I-00133 Rome, Italy  \label{romauniv}
\and
Max Planck Institute for Astronomy, K\"{o}nigstuhl 17, D-69117, Heidelberg, Germany  \label{heidelberg}
\and 
INAF -- Osservatorio di Cagliari, via della Scienza 5, I-09047 Selargius, CA, Italy  \label{cagliari}
\and
Dip. di Fisica e Astronomia Galileo Galilei -- Universit$\grave{\rm a}$ di Padova, Vicolo dell'Osservatorio 2, 35122, Padova, Italy  \label{padovauniv}
%INAF-IAPS Istituto di Astrofisica e Planetologia Spaziali, Via del Fosso del Cavaliere 100, 00133, Roma, Italy \label{iaps}
%\and
%Dipartimento di Fisica, Universit\`{a} degli Studi di Torino, via Pietro Giuria 1, I-10125 Torino, Italy  \label{torinouniv}
%\and
%Thüringer Landessternwarte Tautenburg, Sternwarte 5, 07778, Tautenburg, Germany  \label{tautenburg}
%\and
%Astronomy Department, 96 Foss Hill Drive, Van Vleck Observatory 101, Wesleyan University, Middletown, CT 06459, US  \label{wesleyan}
%\and
%Dipartimento di Fisica e Chimica Emilio Segr{\'e} - Università di Palermo, Piazza del Parlamento, 1, 90134, Palermo, Italy  \label{palermouniv}
%\and
%Centro de Astrobiolog{\'i}a (CSIC-INTA), Carretera de Ajalvir km 4 - 28850 Torrej{\'o}n de Ardoz, Madrid, Spain  \label{inta}
%\and
%Dipartimento di Fisica, Universit\`{a} degli Studi di Milano Bicocca, Piazza dell'Ateneo Nuovo, 1, I-20126 Milano, Italy \label{milanouniv}
%\and
             }
             \offprints{F.~Borsa\\ \email{francesco.borsa@inaf.it}}

   \date{Received ; accepted }

% \abstract{}{}{}{}{} 
% 5 {} token are mandatory
 
  \abstract
   {The detection of lines in emission in planetary atmospheres provides direct evidence of temperature inversion. We confirm the trend of ultra-hot Jupiters orbiting A-type stars showing 
   temperature inversions on their daysides, by detecting metals emission lines in the dayside of \object{KELT-20b}. We first detect the planetary emission by using the G2 stellar mask of the HARPS-N pipeline, which is mainly composed of neutral iron lines, as a template. Using neutral iron templates, we perform a retrieval of the atmospheric temperature-pressure profile of the planet, confirming a thermal inversion.
   Then we create models of planetary emission of different species using the retrieved inverted temperature-pressure profile. By using the cross-correlation technique, we detect \ion{Fe}{i}, \ion{Fe}{ii} and \ion{Cr}{i} at signal-to-noise ratio levels of 7.1, 3.9 and 3.6, respectively.
   The latter is detected for the first time in emission in the atmosphere of an exoplanet. 
 Contrary to \ion{Fe}{i}, \ion{Fe}{ii} and \ion{Cr}{i} are detected only after the occultation and not before, hinting for different atmospheric properties in view on the pre- and post- occultation orbital phases. A further retrieval of the temperature-pressure profile performed independently on the pre- and post- occultation phases, while not highly significant, points to a steeper thermal inversion in the post-occultation.
   }

   \keywords{planetary systems --  techniques: spectroscopic  -- planets and satellites: atmospheres -- stars:individual:KELT-20}
   \maketitle
%
%________________________________________________________________
\section{Introduction\label{sec:intro}}

Ultra-hot Jupiters (UHJs) are highly irradiated gas giant planets with equilibrium temperatures exceeding 2000 K 
and hosting atmospheres presenting substantial H$_2$ dissociation and H$^-$ opacity \citep[e.g.,][]{2018ApJ...855L..30A,2018A&A...617A.110P,2018ApJ...857L..20B}. 
They show different atmospheric properties from the classic hot Jupiters. In particular, one key characteristic is the atmospheric thermal inversion, which appears to happen when the equilibrium temperature reaches $\sim$1700 K, with observational evidence for a transition between the two regimes \citep{2020A&A...639A..36B}. 
\citet{fossati2021} showed that non-LTE effects play a significant role in determining the shape of the temperature inversion.
While previous studies predicted the temperature inversion to be caused mainly by TiO and VO \citep[e.g.][]{2003ApJ...594.1011H,2008ApJ...678.1419F}, more recent works showed that also the presence of atomic species (e.g., Fe, Ti) alone (coupled with the thermal dissociation of infrared coolants like H$_2$O) can cause atmospheric inversions \citep[e.g.][]{2018ApJ...866...27L,2019ApJ...876...69L}. \citet{2019ApJ...876...69L} suggest that UHJs can have a temperature inversion originating from atomic absorption depending on the stellar spectral type.
This presence of atomic species has been observationally confirmed with high-resolution spectroscopy, mainly in transmission \citep[e.g.,][]{2019arXiv190502096H,2020AJ....160..228K,gibson2020,2021A&A...645A..24B}. 
 
 High-resolution spectroscopy allows us to distinguish between the planetary and stellar spectrum, owing to the time resolved Doppler shift due to the planetary orbital motion.
Moreover, it enables us to resolve individual lines, 
ensuring the unambiguous detection of atomic and molecular species.
High-resolution emission spectroscopy can aid in the determination of the primary thermal inversion causing opacity sources.
After the first detection of neutral iron emission from the day-side of a UHJ \citep[KELT-9b,][confirmed in \citet{kasper2021}]{2020ApJ...894L..27P}, neutral iron in emission has since been detected at high resolutions in WASP-33b \citep{2020ApJ...898L..31N} and in WASP-189b \citep{2020A&A...640L...5Y}, two other UHJs orbiting A-type stars.

\object{KELT-20b} \citep{Lund2017}, also known as MASCARA-2b \citep{Talens2018}, is an ultra-hot Jupiter ($T_\mathrm{eq}\sim$2200 K) orbiting a fast rotating (\vsini$\sim$116 \kms) A2-type star in $\sim$3.5 days.
Its atmosphere is the subject of many investigations, with high-resolution transmission spectroscopy leading to the detection of different species like Na, H, Mg, \ion{Fe}{i}, \ion{Fe}{ii}, \ion{Ca}{ii}, Cr 
and possibly FeH \citep[][]{2018A&A...616A.151C,2019A&A...628A...9C,2020A&A...638A..26S,Nugroho2020,2020A&A...641A.120H,2020AJ....160..228K}. 
The \ion{Fe}{i} features have been further used to detect the atmospheric Rossiter-McLaughlin effect \citep{rainer} and to bring evidence of probable atmospheric variability \citep{Nugroho2020,rainer}.

In this work we present the high-resolution detection of multiple-species in emission from the dayside of KELT-20b, suggesting the presence of an atmospheric thermal inversion.
In Sect.~\ref{sec:data_sample} we present our dataset, and in Sect.~\ref{sec:emission} the detection of emission from the planet.
{We then perform a T-P profile retrieval in Sect.~\ref{sec:retrieval}, search for emission from different atomic species in Sect.~\ref{sec:crosscorr}, and conclude with a discussion and final remarks in Sect.~\ref{sec:conclusion}.

%__________________________________________________________________

\section{Data sample\label{sec:data_sample}}

In the framework of the GAPS programme \citep[][]{2019A&A...631A..34B,2020A&A...639A..49G,2021Natur.592..205G}, 
we observed KELT-20 
with the HARPS-N and GIANO-B high-resolution spectrographs, mounted at the Telescopio Nazionale Galileo. We used the GIARPS configuration \citep{giarps}, which allows us to simultaneously use the two spectrographs, obtaining high-resolution spectra in the wavelength range $\sim$390-690 $\rm nm$ and $\sim$940-2420 $\rm nm$.
In this work, we analyse only the HARPS-N spectra, leaving the analysis of the GIANO-B ones to another work. 
We observed both pre- (0.4-0.44) and post- (0.52-0.59) eclipse phases (where 0.0 and 0.5 correspond to the transit and secondary eclipse phases, respectively), with integrations of 600 sec for each spectrum.
While fiber A of the spectrograph was centered on the target, fiber B was monitoring the sky simultaneously. 
The weather conditions on both nights suffered from calima.
A log of the observations is reported in Table~\ref{tab:log}.

\begin{table}
\begin{center}
\caption{KELT-20b HARPS-N observations log.}
\label{tab:log}
\footnotesize
\begin{tabular}{cccccc}
 \hline\hline
 \noalign{\smallskip}
Night \# & Night$^{1}$ & Phase & N$_{\rm obs}$ & Airmass & S/N$_{\rm ave}$\\
 \noalign{\smallskip}
 \hline
\noalign{\smallskip}
1 & 11 Oct 2020  & 0.53-0.59  & 30 &1.0-2.5 &165 \\
2 & 14 Oct 2020  & 0.40-0.44 & 25 & 1.1-2.0 &200 \\
\noalign{\smallskip}
 \hline
   \multicolumn{3}{l}{$^{1}$\footnotesize{Start of night civil date.}} \\
\end{tabular}
\end{center}
\end{table}

%______________________________________________________________
\section{Planetary emission detection\label{sec:emission}}

In order to check if the planetary Doppler trail was present (Fig. \ref{fig:tomo}), 
we started our analysis from the stellar cross-correlation functions (CCFs) extracted by version 3.7 of the HARPS-N Data Reduction Software (DRS) pipeline \citep{2014SPIE.9147E..8CC}, using the YABI interface with custom parameters \citep[e.g.,][]{borsa}. In particular, we used a G2 stellar mask (the mask already rejects regions strongly contaminated by tellurics) and enlarged the CCF width to the range [-300:300] \kms. This was done to take into account the large \vsini\ of the star and to sample the velocity range of the possible planetary signal, which is expected to appear offset from the stellar velocity by dozens of \kms\ at the observed orbital phases.
The stellar mask used to compute the CCF is mainly composed of \ion{Fe}{i} lines \citep[][]{espresso1}, 
and is thus a good template for searching for neutral iron in the planetary atmosphere.

\begin{figure*}%[!ht]
\centering
\includegraphics[width=\linewidth]{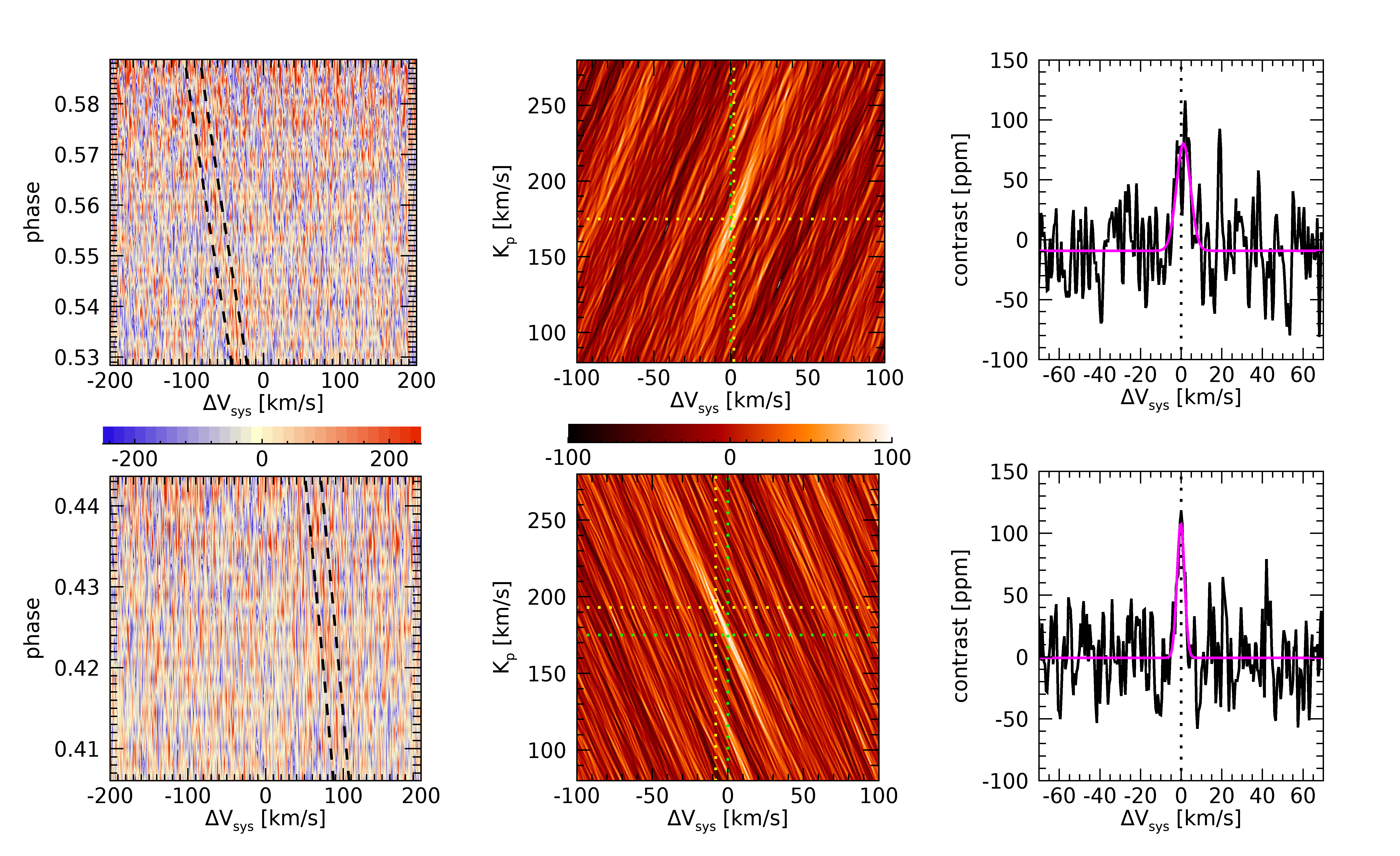}
\caption{The first row refers to night 1 and the second row to night 2. {\it(Left panels)}: 2D tomography of the CCF residuals maps. The dashed lines bracket the expected planetary signal position.
{\it(Central panels)}: K$_{\rm p}$-$\Delta$V$_{\rm sys}$ maps. The green dotted lines mark the theoretical planetary position.
{\it(Right panels)}: Planetary emission signal averaged at the theoretical K$_{\rm p}$. The magenta line represents the Gaussian fit to the data.
}
\label{fig:tomo}
\end{figure*}

We analysed each of the two nights of observations (Table \ref{tab:log}) independently. We first shift the CCFs into the stellar rest frame using the Keplerian orbital solution calculated with the parameters of Table~\ref{table:system}. 
The CCFs suffered from an evident slope, due to the imperfect color correction caused by the spectral type of the star, which does not have a proper color correction template.
We thus normalised the CCFs by dividing for a linear fit performed excluding the range [-140:140] \kms\ (i.e. we fit only the continuum). 
Then we created a master CCF (CCF$_{\rm master}$) by doing a median of all the CCFs, and divided all the CCFs for this CCF$_{\rm master}$. 

This normalisation allowed us to null the largely stationary stellar spectrum, but not the planetary one, which Doppler shifts across numerous different pixels over the course of the observing sequence.
The median approach has been successfully used in other works involving planetary emission \citep{2020ApJ...894L..27P} and reflectance spectroscopy \citep{2020arXiv201210435S}.
At this stage in the analysis process, the planetary Doppler trail within the CCF residuals was weakly present (Fig. \ref{fig:tomo}, left panels).

For a range of K$_{\rm p}$ values from 0 to 300 \kms, in steps of 1 \kms, we averaged the CCF$_{\rm res}$ after shifting them in the planetary rest frame, which was done by subtracting the planetary radial velocity computed for each exposure as $v_{p}=K_{\rm p} \times \sin{2\pi \phi}$, with $\phi$ the orbital phase. 
We thus created the K$_{\rm p}$-$\Delta$V$_{\rm sys}$ maps (with $\Delta$V$_{\rm sys}$ the difference from the literature reported systemic velocity V$_{\rm sys}$), that we used to check the presence of a significant signal close to the expected planetary K$_{\rm p}$ (Fig. \ref{fig:tomo}, central panels).
We then looked at the signal at the nominal K$_{\rm p}$ (i.e., at the row of the K$_{\rm p}$-$\Delta$V$_{\rm sys}$ map corresponding to K$_{\rm p}$=175 \kms), and performed a 1D Gaussian fit to measure its amplitude (Fig.~\ref{fig:tomo}, right panels). 

There is a significant detection of planetary emission for both nights, with signal-to-noise (S/N) levels of 3.2 and 4.0 for night 1 and night 2, respectively. The S/N of the detections is calculated with respect to the noise level of the continuum ($\sim$27-28 ppm on both nights), which was evaluated by calculating the standard deviation of the K$_{\rm p}$-$\Delta$V$_{\rm sys}$ maps far from where any stellar or planetary signal is expected.
The planetary signal is detected in emission (i.e., opposite to the stellar signal, which is in absorption), which is an unambiguous sign of thermal inversion in its atmosphere \citep{2020ApJ...894L..27P}. Without a thermal inversion, the signal would have been seen in absorption (i.e., a negative signal).
The results of the Gaussian fits are shown in Table~\ref{tab:numbers}, expressed as the ratio between planetary and stellar flux.
The amplitudes of the planetary signal at the theoretical K$_{\rm p}$ are consistent for both nights, but there is difference in their FWHMs.
We note that pre- and post-occultation orbital phases show different regions of the planetary atmosphere, and any difference because of non-identical temperature, 
chemical composition and abundances is indeed possible. By using the theoretical K$_{\rm p}$ for two different nights and different orbital phases, we intrinsically assume a circular Keplerian orbit and that there is no atmospheric dynamics influencing the observed velocities.
This latter point, in particular, has been questioned by recent results on this planet, which show possible variability of the K$_{\rm p}$ detected in transmission with an iron mask \citep{Nugroho2020,rainer}.
Comparing in detail the results of different nights taken at different orbital phases to constrain the three dimensional characteristics of a planetary atmosphere
requires a more complex three-dimensional modeling analysis framework. Such a framework is out of the scope of this manuscript and will be the subject of another work (Pino et al. in prep.). 
 
We then combined the K$_{\rm p}$-$\Delta$V$_{\rm sys}$ maps from each night (Fig.~\ref{fig:2nights}) by summing them and fit the signal resulting from the combination with a 2D-Gaussian, obtaining values of K$_{\rm p}$=173$\pm$9 \kms and $\Delta$V$_{\rm sys}$=0.6$\pm$3.6 \kms.

\begin{figure}%[!ht]
\centering
\includegraphics[width=\linewidth]{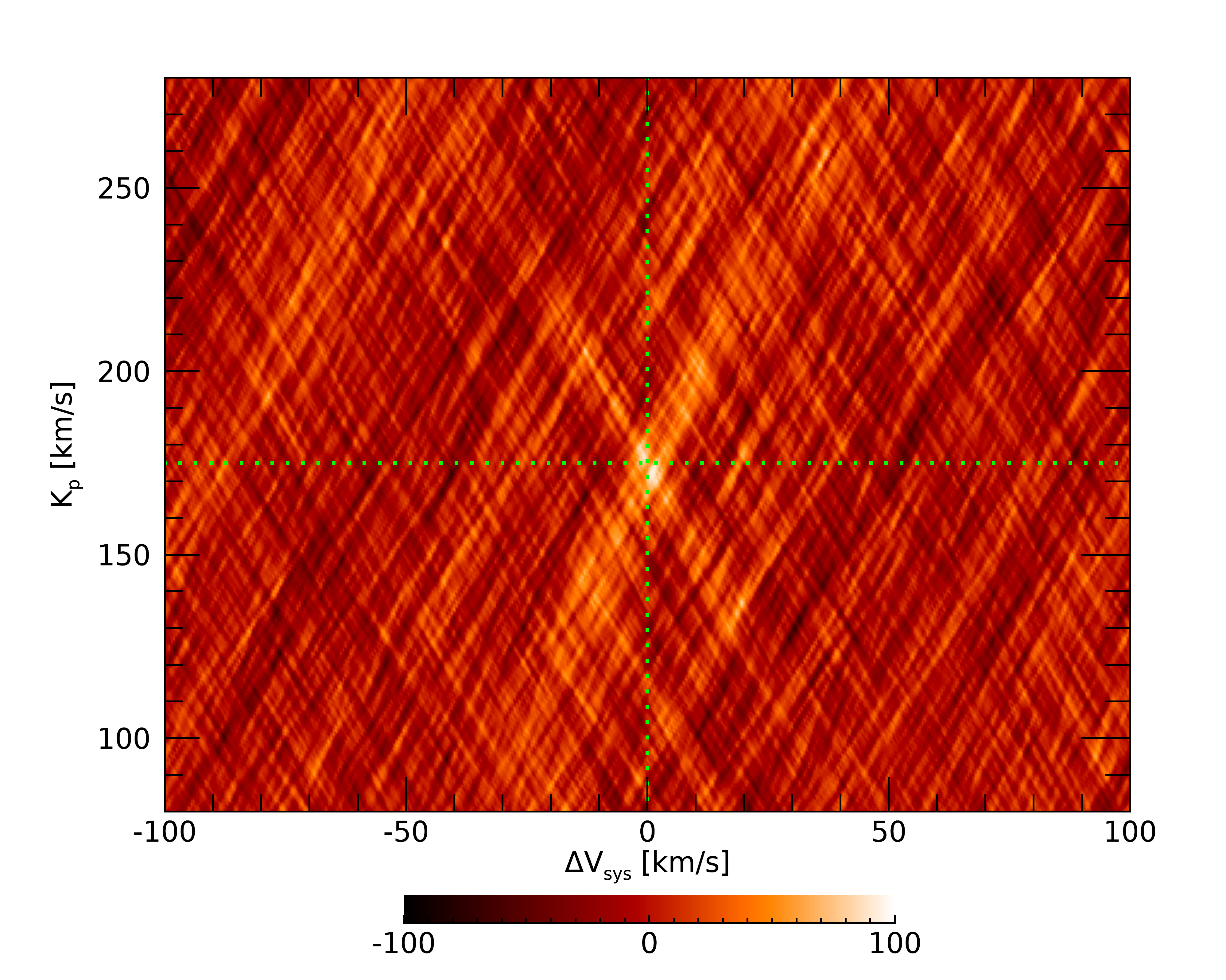}
\caption{K$_{\rm p}$-$\Delta$V$_{\rm sys}$ map of the two nights combined. The green dotted lines mark the expected planetary position at K$_{\rm p}$=175 \kms.
The colour scale is in ppm.
}
\label{fig:2nights}
\end{figure}

\begin{table}
\begin{center}
\caption{KELT-20b emission detections with the stellar G2 mask as a template.}
\label{tab:numbers}
\footnotesize
\begin{tabular}{cccc}
 \hline\hline
 \noalign{\smallskip}
Night & contrast [ppm] & center [\kms] & FWHM [\kms] \\
 \noalign{\smallskip}
 \hline
\noalign{\smallskip}
1 & 89$\pm$10  & 1.1$\pm$ 0.5  & 8.1$\pm$1.1   \\
2 & 108$\pm$12   & -0.2$\pm$ 0.2 &  4.3 $\pm$0.7 \\
\noalign{\smallskip}
 \hline
\end{tabular}
\end{center}
\end{table}

%______________________________________________________________
\section{T-P profile retrieval \label{sec:retrieval}}

In order to constrain the T-P profile, we performed an atmospheric retrieval following a framework similar to that presented in \citet{2020A&A...640L...5Y}, using the same 
likelihood function as presented in \citet{gibson2020}.

After normalising the stellar spectra, we corrected for tellurics by exploiting the relation between their depth and airmass \citep[e.g.,][]{2008A&A...487..357S,2010A&A...523A..57V}. 
Then we created a master stellar spectrum by shifting the spectra in the stellar rest frame and taking a median spectrum. 
The orbital velocity of the planet coupled with the instrumental resolution ensures that the planetary signal was removed from 
this master spectrum \citep[e.g.,][]{2020arXiv201210435S}.
All the spectra were then normalised for this master stellar spectrum, and all the residual spectra $S_{res}$ were shifted in the planetary rest frame using the theoretical K$_{\rm p}$ value, that we assume fixed. 
This K$_{\rm p}$ value was also confirmed by our analysis of Sect.~\ref{sec:emission}.
The residual spectra of both nights were then merged together with a weighted average, creating the final planetary spectrum $R_i$.
The error-bar on the flux of each wavelength point $\sigma_i$ was calculated on the stellar spectrum as the square root of the flux, and then propagated through the analysis.

\

The detection of the stellar CCF in emission at the planetary K$_{\rm p}$ proved the presence of neutral iron in the planet (Sect.~\ref{sec:emission}). 
We then performed the retrieval by generating neutral iron emission model spectra for the planetary atmosphere
with \texttt{petitRADTRANS} \citep{pRT}, assuming solar metallicity and equilibrium chemistry.
Following \citet{2020A&A...640L...5Y}, we considered a two-points T-P profile (Fig. \ref{fig:bestTP}). The temperature in the atmosphere below the higher pressure point and beyond the lower pressure point was considered isothermal, and the slope between the two points was defined by a gradient

\begin{equation}
T_{\rm slope}= \frac{T_2- T_1}{\log_{10}P_2- \log_{10}P_1} 
\label{equazTP}
\end{equation}

with the temperature changing linearly with $\log_{10}P$.
The generated model spectra were then divided for a stellar blackbody spectrum with T$_{\rm eff}$=8980 K (Table \ref{table:system}) to have the model in units F$_{\rm p}$/F$_{\rm s}$, convolved with the instrumental profile (using a resolving power of 115,000) and continuum normalised.
When using a stellar blackbody spectrum we ignored the dependence on the stellar lines.

\

For the T-P profile retrieval, we used the wavelength range 4000-6500 \AA, discarding the low S/N bluer wavelengths and the heavily telluric affected red ones.
The retrieval was performed in a Bayesian framework by employing a differential evolution 
Markov chain Monte Carlo (DE-MCMC) technique \citep{TerBraak2006, Eastmanetal2013}, running 10 DE-MCMC chains. 
We left as free parameters $\log_{10}P_1$, T$_1$, $\log_{10}P_2$, T$_2$ and $\Delta$V$_{\rm sys}$, for which we set uninformative priors. 
For the identification of the burn-in steps and the convergence and well-mixing of the DE-MCMC chains, we followed the same criteria as in \citet{Eastmanetal2013}.
We took the medians and the 15.86\% and 84.14\% quantiles of the posterior distributions as the best values and $1\sigma$ uncertainties of the model parameters. 
By parallelising our retrieval code, we were able to reduce the computing time and to generate models in continuum, without using a grid like in \citet{2020A&A...640L...5Y}. This mean that a new model of iron in the atmosphere was computed at each step of each chain.

The best retrieved values are shown in Table~\ref{tab:bestTP}, with the best T-P profile from the retrieval shown in Fig.~\ref{fig:bestTP} and the corner plot with the MCMC posterior distributions in Fig.~\ref{fig:bestTP_corner}.

\begin{figure}%[!ht]
\centering
\includegraphics[width=\linewidth]{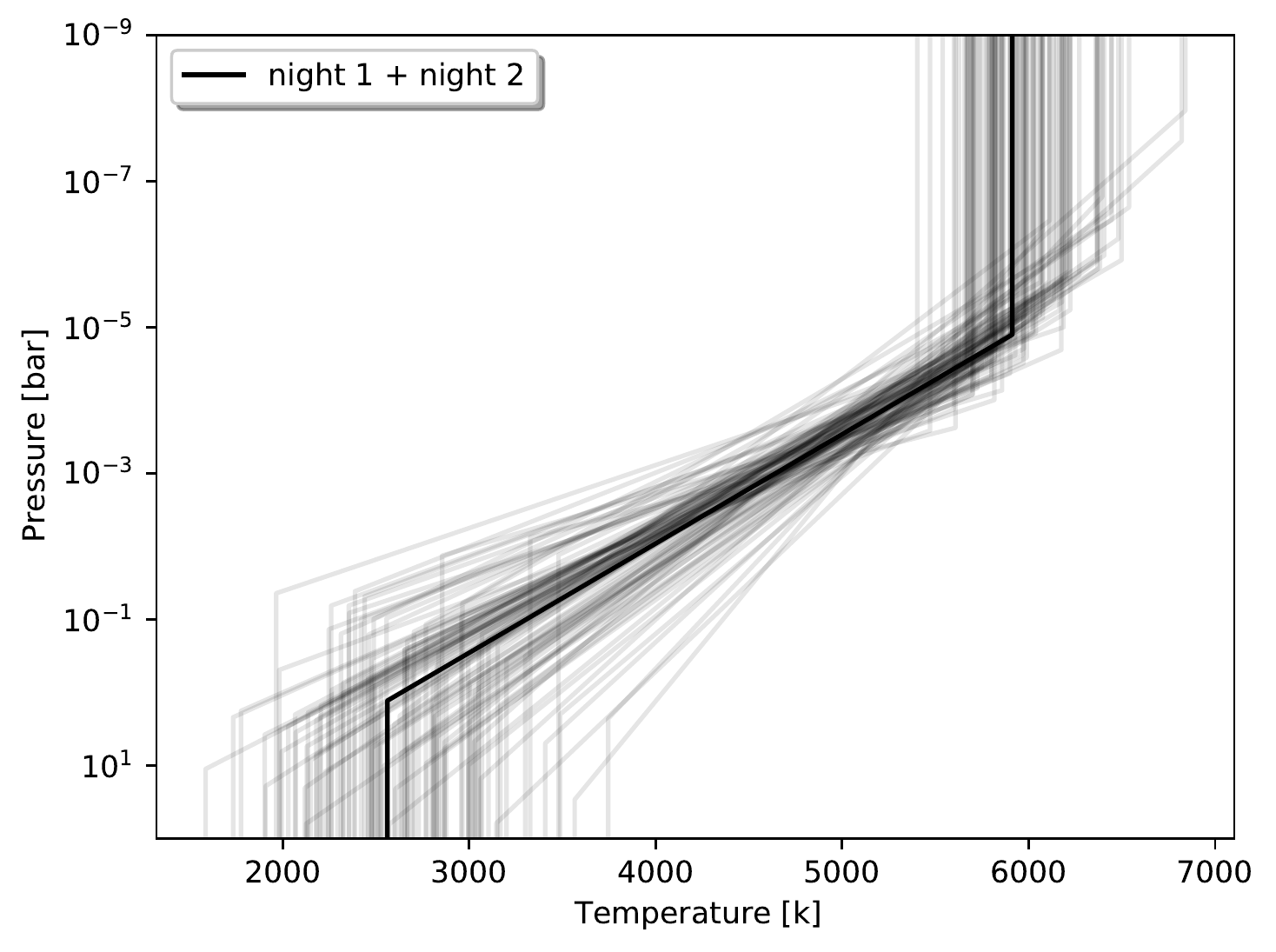}
\caption{T-P profile retrieved for the dayside of KELT-20b in this work. Thinner lines represent 100 T-P profiles randomly extracted from the MCMC posteriors.
}
\label{fig:bestTP}
\end{figure}

\begin{figure}%[!ht]
\centering
\includegraphics[width=\linewidth]{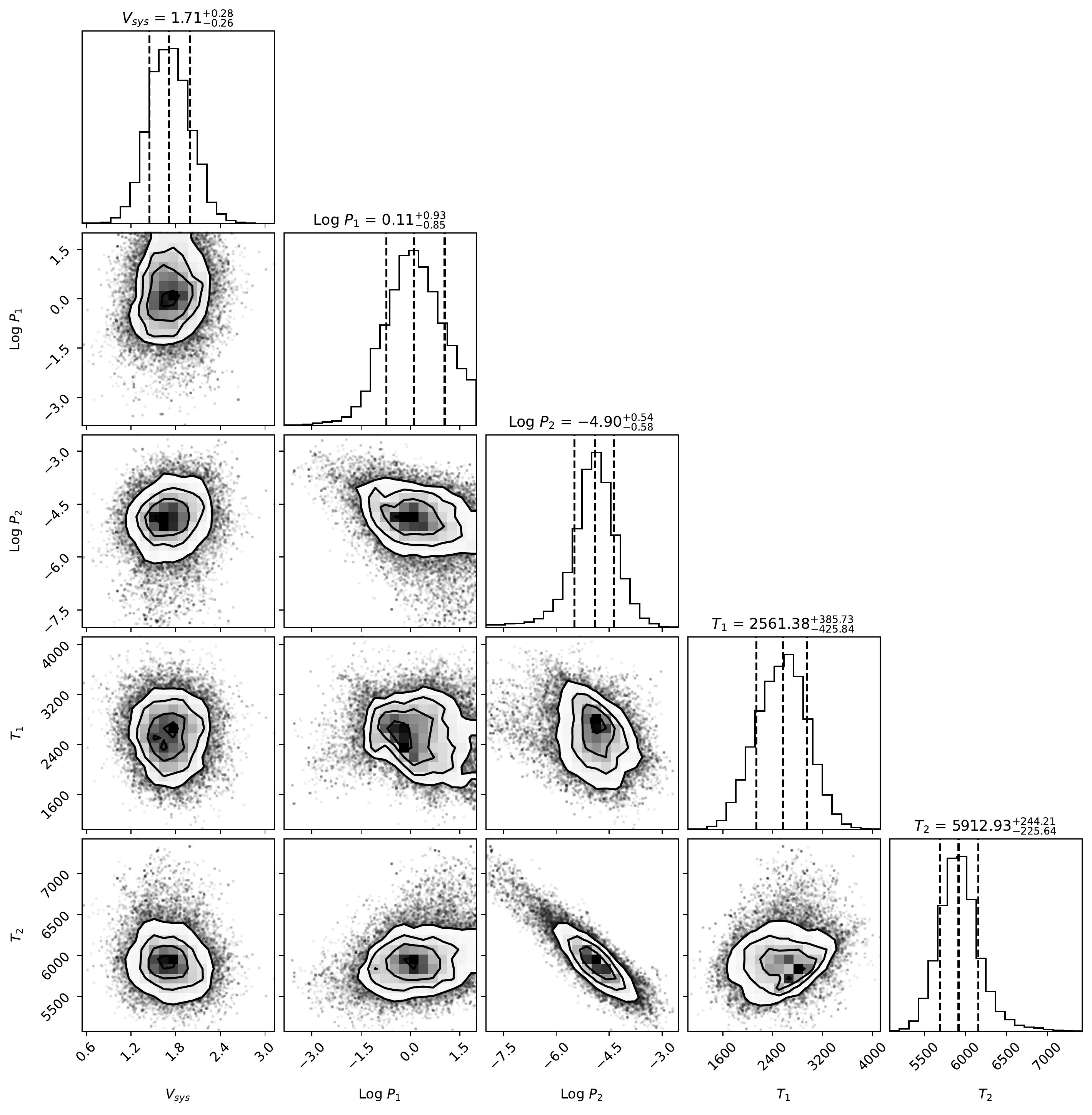}
\caption{Corner plot representing the posterior distribution of variables used for the DE-MCMC computations of the T-P profile parameters.}
\label{fig:bestTP_corner}
\end{figure}

\begin{table}
\begin{center}
\caption{Results of the T-P profile retrieval.}
\label{tab:bestTP}
\footnotesize
\begin{tabular}{ccc}
 \hline\hline
 \noalign{\smallskip}
Parameter & Value & Unit \\
 \noalign{\smallskip}
 \hline
\noalign{\smallskip}
$\Delta$V$_{\rm sys}$  &$1.71_{-0.26}^{+0.28}$  & \kms \\
\noalign{\smallskip}
$\log_{10}P_1$ & $-0.11_{-0.85}^{+0.93}$  & log bar   \\
\noalign{\smallskip}
T$_1$ & 2561.38$_{-425.84}^{+385.73}$ & K\\
\noalign{\smallskip}
$\log_{10}P_2$ & -4.90$_{-0.58}^{+0.54}$   & log bar \\
\noalign{\smallskip}
T$_2$ & 5912.93$_{-225.64}^{+244.21}$ & K\\
\noalign{\smallskip}
 \hline
\end{tabular}
\end{center}
\end{table}

%______________________________________________________________
\section{Searching for atomic species \label{sec:crosscorr}}

Once determined the T-P profile using \ion{Fe}{i} models, we checked for the presence of emission from other species.
We looked first at \ion{Fe}{i}, the same used in the retrieval, to confirm that this is the main source of the detection with the stellar mask. Then we looked also for \ion{Fe}{ii}, \ion{Ti}{i} and \ion{Cr}{i}, which are among the other principal components of the stellar mask \citep[e.g.,][]{espresso1}. 

Emission model spectra for the planetary atmosphere were generated in the same way as in Sect. \ref{sec:retrieval}, using the retrieved T-P profile.

The \ion{Fe}{i}, \ion{Ti}{i} and \ion{Cr}{i} models present many emission lines, thus making them sensitive to the cross-correlation 
method (Fig. \ref{fig:models}). This is not the case for 
the \ion{Fe}{ii} model, which however shows fewer but stronger lines.

\begin{figure}[!ht]
\centering
\includegraphics[width=\linewidth]{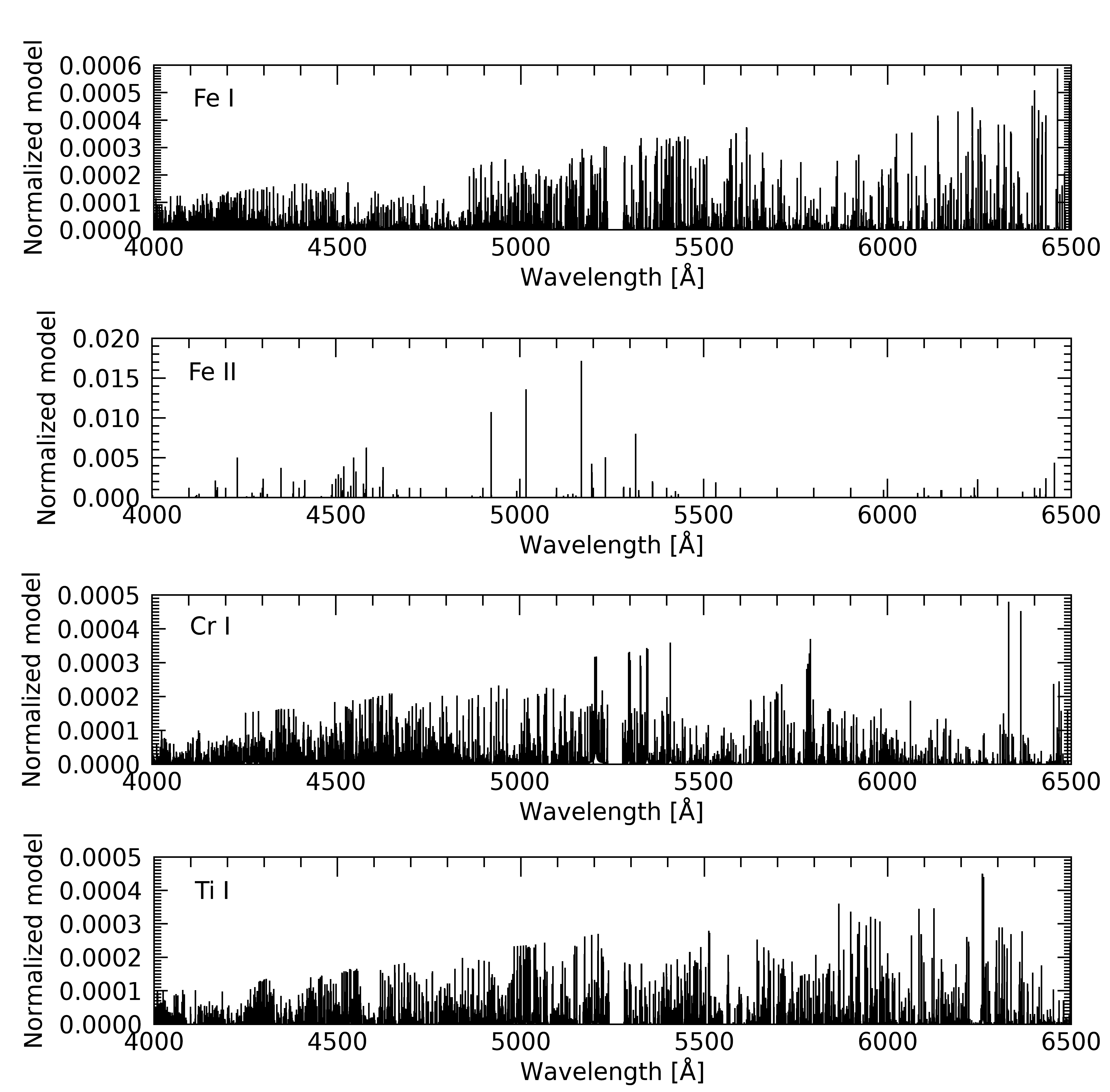}
\caption{From top to bottom, the models of \ion{Fe}{i}, \ion{Fe}{ii}, \ion{Cr}{i} and \ion{Ti}{i} used for the cross-correlations. The models are normalised to unity.
}
\label{fig:models}
\end{figure}

The cross-correlation was performed on the residual spectra $S_{res}$ (see Sect. \ref{sec:retrieval}) as in \citet{2021A&A...645A..24B}, normalising the model to unity and thus preserving the flux information \citep[e.g.,][]{2019arXiv190502096H}.
In our analysis, we zero out any lines that have strengths less than 1\% of the maximum line strength within the modeled wavelength range \citep[e.g.,][]{2019arXiv190502096H}. 
We selected a step of 1 \kms\  and a velocity range [-300, 300] \kms. 
The spectra were divided into segments of 200 $\AA$ \citep[e.g.,][]{2019arXiv190502096H}, then the cross-correlation was performed for each segment covering the range 3900-6500 $\AA$. 
We performed a 5$\sigma$-clipping on the processed data frames to eliminate outliers and masked the wavelength range 5240-5280 $\AA$ (i.e., the most affected by telluric contamination).
 Then for each exposure, we applied a weighted average of the cross-correlations of the single segments, where the weights applied to each 
 segment were the square of the inverse of its standard deviation and the depths of the lines in the model.

For a range of K$_{\rm p}$ values from 0 to 300 \kms, in steps of 1 \kms, we averaged all the cross-correlation functions after shifting them in the planetary restframe. 
This was done by subtracting the planetary radial velocity calculated for each spectrum as $v_{p}=K_{\rm p} \times \sin{2\pi \phi}$, with $\phi$ the orbital phase. 
We thus created the K$_{\rm p}$ versus $\Delta$V$_{\rm sys}$ maps, 
to check that the signal is found close to the planet radial-velocity semi-amplitude and stellar systemic velocity, as is expected if the signal is of planetary origin.
We then evaluated the noise by calculating the standard deviation of the K$_{\rm p}$ versus $\Delta$V$_{\rm sys}$ maps far from where any stellar or planetary signal is expected.
The S/N of the detections (Table \ref{tab:results}) was calculated by dividing the best K$_{\rm p}$ cross-correlation function by the noise, and by fitting a Gaussian function to the result (see Fig. \ref{fig:contours} for the detections of night 1).

The results indicate the detection of \ion{Fe}{i} in both nights, with S/N of 5.3 and 6.0, respectively. On the first night, we also detect \ion{Fe}{ii} with S/N=3.9 and \ion{Cr}{i} with S/N=3.6. These two elements are not detected in night 2, while the level of noise is comparable to that of night 1. We do not detect \ion{Ti}{i}. The results of the cross-correlations in individual nights and combining the two nights are reported in Table \ref{tab:results}. 
We note that when combining the two nights, the S/N of the detections of \ion{Fe}{ii} and \ion{Cr}{i} remains almost the same. This gives us further confidence in the reliability of the planetary signal, as also without detection when adding the second night this does not bring a strong destructive interference.

\begin{table*}
\begin{center}
\caption{Results of the cross-correlation with theoretical models. ND stands for no detection.}
\label{tab:results}
\footnotesize
\begin{tabular}{c|ccc|ccc|c}
 \hline\hline
 \noalign{\smallskip}
 &  \multicolumn{3}{c|}{night 1} & \multicolumn{3}{c|}{night 2} & combined\\
Element & K$_{\rm p}$ [\kms] & $\Delta$V$_{\rm sys}$ [\kms] & S/N  & K$_{\rm p}$ [\kms] & $\Delta$V$_{\rm sys}$ [\kms] & S/N & S/N\\
 \noalign{\smallskip}
 \hline
\noalign{\smallskip}
\ion{Fe}{i}  & 184$^{+5}_{-10}$& 5$^{+3}_{-2}$ & 5.3 &  176$^{+3}_{-3}$ & 1$^{+2}_{-2}$ & 6.0 & 7.1 \\
\noalign{\smallskip}
\ion{Fe}{ii} & 159$^{+5}_{-4}$ & -4$^{+2}_{-2}$ & 3.9 &  ND & ND & ND & 3.3\\
\noalign{\smallskip}
\ion{Cr}{i} & 158$^{+6}_{-5}$ & -1$^{+2}_{-3}$ & 3.6 &  ND & ND & ND & 3.5 \\
\noalign{\smallskip}
 \hline
\end{tabular}
\end{center}
\end{table*}

\begin{table}
\begin{center}
\caption{Significance of the detections in night 1 calculated with different statistical methods.}
\label{tab:signi}
\footnotesize
\begin{tabular}{ccccc}
 \hline\hline
 \noalign{\smallskip}
Element & S/N & T-test & KS & bootstrap\\
 \noalign{\smallskip}
 \hline
\noalign{\smallskip}
\ion{Fe}{i}  & 5.3 & 6.8$\sigma$ & 5.4$\sigma$ & 5.9$\sigma$ \\
\ion{Fe}{ii} & 3.9 & 3.2$\sigma$ & 3.1$\sigma$ & 3.6$\sigma$\\
\ion{Cr}{i} & 3.6 & 3.9$\sigma$ & 3.1$\sigma$ & 3.1$\sigma$\\
\noalign{\smallskip}
 \hline
\end{tabular}
\end{center}
\end{table}

\begin{figure}%[!ht]
\centering
\includegraphics[width=\linewidth]{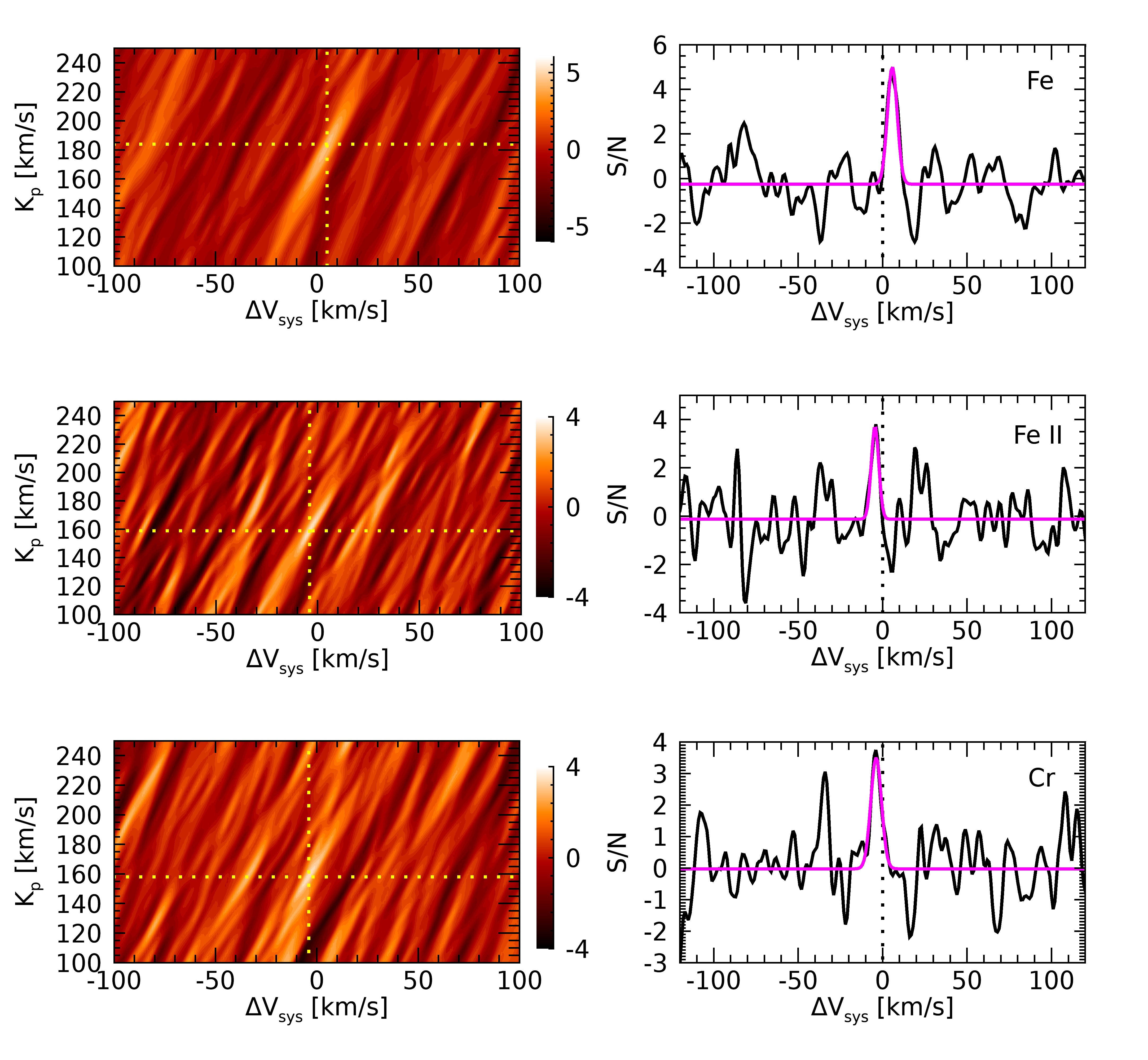}
\caption{From top to bottom, detections of \ion{Fe}{i}, \ion{Fe}{ii}, and \ion{Cr}{i} in night 1.
              Left panels show the K$_{\rm p}$-$\Delta$V$_{\rm sys}$ maps, with the colour scale in S/N. The yellow dotted lines represent the best signal position.
              Right panels show the signal at the best K$_{\rm p}$ position, with a Gaussian fit shown in magenta.}
\label{fig:contours}
\end{figure}

To assess the significance of our detections that we got only in one night, we performed also other statistical tests on this night only.
The first one is performed by comparing the in-trail versus out-of-trail samples of the cross-correlation functions and performing a Welch 
t-test \citep[e.g.,][]{2017AJ....153..138B,2017AJ....154..221N,2018A&A...615A..16B,2019A&A...625A.107G}.
To avoid oversampling, we resampled the cross-correlation functions to 1.5 \kms, which is about twice the average pixel size of the instrument \citep[$\sim$0.8 \kms,][]{2014SPIE.9147E..8CC} and close to its half-width at half maximum ($\sim$1.3 \kms). For each species tested, we selected the interval 20<|v|<100 \kms\ for the out-of-trail sample, while for the in-trail one we optimized the width choosing the one that maximises the detection. We observed in any case a low dependence of the final significance on the width of the in-trail sample.
The error-bars on the two samples were calculated as the square root of the number of data points in each bin (Fig. \ref{fig:signi_FeI}).
This test gives a confidence of 6.8$\sigma$, 3.2$\sigma$ and 3.9$\sigma$ for \ion{Fe}{i}, \ion{Fe}{ii}, and \ion{Cr}{i}, respectively, that the in-trail and out-of-trail 
distributions are drawn from different parent distributions in night 1. 
The distributions are compatible with Gaussian functions, calculated by assuming 
the average of the samples and their standard deviation as the center and the width of the Gaussian, respectively
 (Fig. \ref{fig:signi_FeI}), thus validating the use of this statistics. 
We note however that \citet{2019MNRAS.482.4422C} suggest that the Welch t-test 
could sometimes overestimate the confidence of detections.
We thus made also Kolmogorov-Smirnov (KS) statistics on the same distributions, obtaining a confidence of 5.4$\sigma$, 3.1$\sigma$ 
and 3.1$\sigma$ for \ion{Fe}{i}, \ion{Fe}{ii}, and \ion{Cr}{i}, respectively, that the distributions in- and out-of-trail are different (Table \ref{tab:signi}).

We did a further statistic significance test by performing a bootstrap, adapting the method proposed in appendix C.2 of \citet{2020A&A...641A.123H} 
to our dataset, where there is no transit.
For each species we shifted each cross-correlation function of the time-series to a random radial velocity taken from a uniform distribution, 
masking all the zones where the planetary signal is expected, and averaged these CCFs.
We then fitted a Gaussian profile with a fixed width of 5, 10 or 20 \kms, centered at a random position in the averaged CCF. 
This was repeated 100,000 times for each species.
In Fig. \ref{fig:bootstrap} we show the resulting distributions coming from random fluctuations for night 1, together with the contrast of the detections, which are 
significantly stronger. We fitted a Gaussian function to each of the random distributions, took the maximum width $\sigma$ of the three (which is always the one created with a fixed width of 5 \kms), and estimated the significance of detection by taking the ratio between the amplitude of the signal and $\sigma$ (Table \ref{tab:signi}).
We confirm the detections of \ion{Fe}{i}, \ion{Fe}{ii} and \ion{Cr}{i} in night 1 with all the statistical methods used, thus validating their robustness.

Following the detection of multiple species, we further investigated the presence of \ion{V}{i}, \ion{V}{ii}, \ion{Y}{i}, \ion{Ca}{i}, \ion{Mg}{i}, AlO, TiO, VO, but without finding statistical significance for their presence in the planetary atmosphere.

%______________________________________________________________
\section{Discussion and conclusions\label{sec:conclusion}}

Our result confirms the tendency of a temperature inversion in the atmosphere of UHJ caused by absorption of UV and optical stellar light by metals. 
KELT-20b is the fourth UHJ for which day-side neutral iron emission is detected with high-resolution spectroscopy. All these planets orbit A-type stars.
\citet{2019ApJ...876...69L} theoretically showed that the slope of the T-P and the temperature range across the temperature inversion in UHJs both increase as the host star effective temperature increases. 
It is thus probably not a casualty that all the detections of temperature inversion with high-resolution spectroscopy up to now are for UHJs orbiting A-type stars.
These stars have strong UV emission, which is the wavelength range where the metals causing temperature inversions present the largest number of lines \citep{fossati2021}.

We added \ion{Cr}{i} to the species discovered in emission from the dayside of an exoplanet, by performing a multi-species atomic detection of planetary emission.
We note that we detected \ion{Fe}{ii} and \ion{Cr}{i} only after occultation, while we see \ion{Fe}{i} better before the occultation. 
In principle, the S/N is almost the same for both nights, so this could hint to different atmospheric properties in the atmosphere at sight during the different orbital phases.
An offset in thermal phase curves has been often observed for short-period exoplanets \citep[e.g.,][and references therein]{2020NatAs...4..453D}, so one possibility is that the temperature (or temperature gradient) is higher in the atmosphere in view after the occultation, making neutral iron to become more ionized and chrome emission visible.

We tried to verify this by performing a further retrieval of the T-P profile in the same way as in Sect. \ref{sec:retrieval}, but now separately before and after occultation.
While the differences are not highly significant, the results point to a steeper thermal inversion for night 1, after the occultation, where we have detections of multiple species (Fig. \ref{fig:TPcomparison}, \ref{fig:N1}, \ref{fig:N2}, Table \ref{tab:bestTP_singlenights}).
It must be noted that our assumption of the T-P profile might be oversimplified and could not represent the true conditions of the atmosphere.

\begin{figure}%[!ht]
\centering
\includegraphics[width=\linewidth]{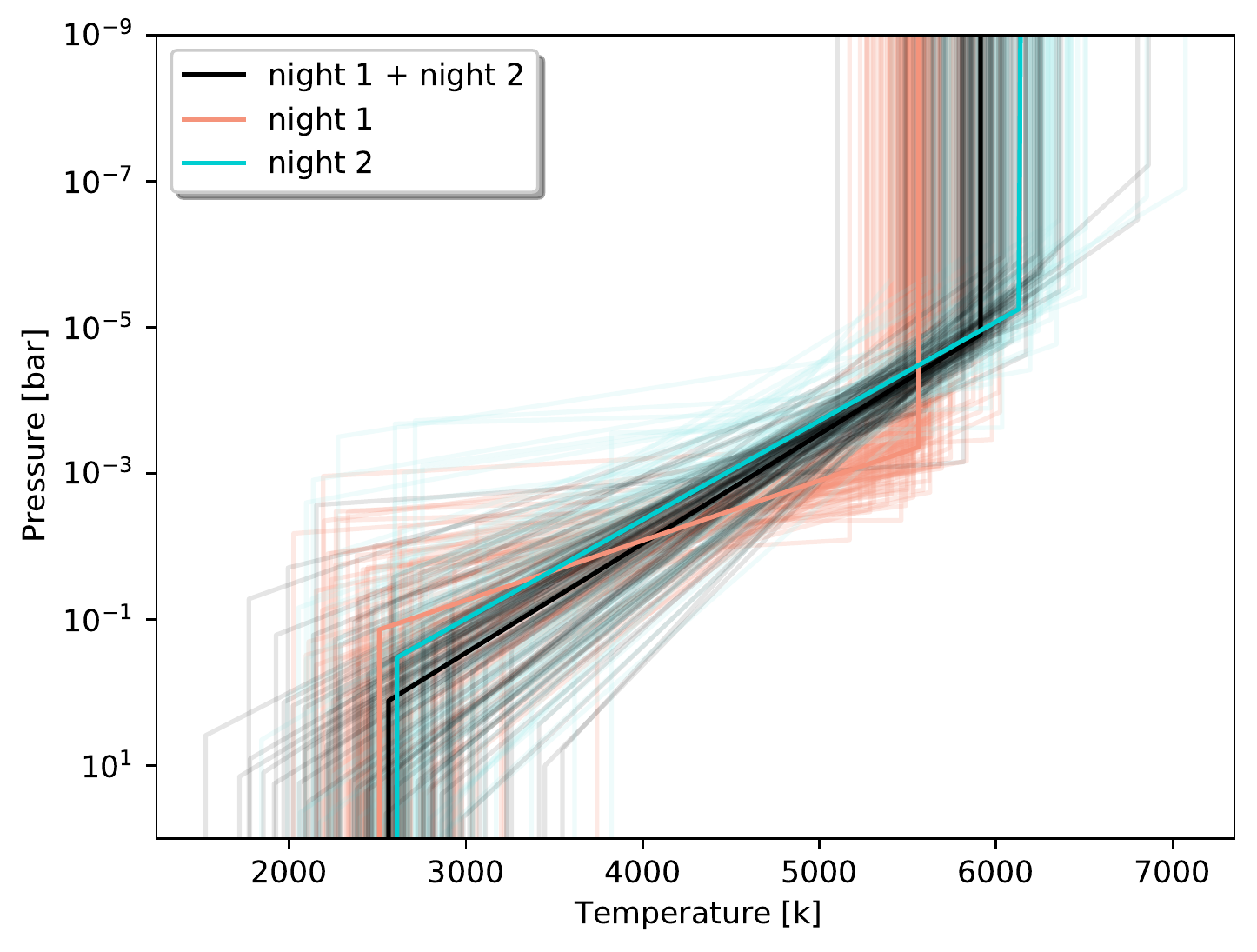}
\caption{T-P profile retrieved for nights 1 and 2. Thinner lines represent 100 T-P profiles randomly extracted from the MCMC posteriors.
}
\label{fig:TPcomparison}
\end{figure}

\begin{table}
\begin{center}
\caption{Results of the T-P profile retrieval for the single nights.}
\label{tab:bestTP_singlenights}
\footnotesize
\begin{tabular}{cccc}
 \hline\hline
 \noalign{\smallskip}
Parameter & Night 1 & Night 2 & Unit \\
 \noalign{\smallskip}
 \hline
\noalign{\smallskip}
$\Delta$V$_{\rm sys}$  &$2.80_{-0.41}^{+0.45}$ & $1.40_{-0.28}^{+0.29}$ & \kms \\
\noalign{\smallskip}
$\log_{10}P_1$ & -0.86$_{-0.96}^{+1.36}$ &  -0.11$_{-1.34}^{+1.39}$ & log bar   \\
\noalign{\smallskip}
T$_1$ & 2508.67$_{-282.96}^{+278.96}$ & 2610.47$_{-303.27}^{+229.20}$ & K\\
\noalign{\smallskip}
$\log_{10}P_2$ & -3.36$_{-0.60}^{+0.47}$ & -5.25$_{-0.74}^{+0.67}$ & log bar \\
\noalign{\smallskip}
T$_2$ & 5561.58$_{-149.86}^{+195.52}$ & 6139.38$_{-287.50}^{+250.69}$ & K\\
\noalign{\smallskip}
 \hline
\end{tabular}
\end{center}
\end{table}

The emission at different planetary phases could be used to constrain the atmospheric dynamics of the planet \citep[e.g.][]{2021A&A...646A..94A},
but an in-depth analysis of different planetary phases however requires a framework of analysis that includes 3D atmospheric modeling.
To verify the real atmospheric origin of the differences found, it will be useful to collect more data with high S/N to check the repeatability of the behaviour and possibly give a stronger constrain on the T-P profiles directly from the observations. 
After the submission of our work, \ion{Si}{i} dayside emission has been detected from the planet \citep[][]{2022A&A...657L...2C}, and a strong temperature inversion consistent with our results was retrieved in its atmosphere using \ion{Fe}{i} templates \citep[][]{2022arXiv220108759Y}. Temperature inversion was retrieved also detecting H$_2$O and CO emission features at low resolution \citep[][]{2022ApJ...925L...3F}.
\ion{Fe}{ii} and \ion{Cr}{i} are firstly detected in this work.

This work demonstrates that high-resolution emission spectroscopy can be exploited much like high-resolution transmission spectroscopy for multiple-species detection in exoplanetary atmospheres, allowing us to investigate the dynamical, chemical, and radiative processes operating in UHJ atmospheres.

%______________________________________________________________

\begin{acknowledgements}
We thank the referee for the useful comments that helped improving the work.
We acknowledge the support by INAF/Frontiera through the "Progetti Premiali" funding scheme of the Italian Ministry of Education, 
University, and Research and from PRIN INAF 2019.
\end{acknowledgements}

%______________________________________________________________

\begin{appendix}

\section{Parameters\label{sec:parameters}}

\begin{table}[!ht]
\caption{Physical and orbital parameters of the KELT-20 system used in this work.}             
\label{table:system}      % is used to refer this table in the text
\centering                          % used for centering table
\begin{tabular}{c c c}        % centered columns (4 columns)
\hline\hline                 % inserts double horizontal lines
 \noalign{\smallskip}
Parameter &  Value & Reference\\    % table heading 
 \noalign{\smallskip}
\hline    
 \noalign{\smallskip}
& Stellar Parameters   & \\
 \noalign{\smallskip}
\hline
 \noalign{\smallskip}
   T$_{\rm eff}$  &  $8980^{+90}_{-130}$~K &1 \\
    \vsini\  & $116.23\pm1.25$~\kms &2\\
   $M_{\star}$  & $1.89^{+0.06}_{-0.05}~M_{\sun}$ &1 \\ 
    $R_\star$  & $1.60 \pm 0.06~R_\sun$ &1\\
 \noalign{\smallskip}
\hline
 \noalign{\smallskip}
& Planetary parameters   & \\
 \noalign{\smallskip}
\hline
 \noalign{\smallskip}
    $M_{\rm p}$ &  $<~3.51~M_{\mathrm{Jup}}$ &3 \\ 
    $R_{\rm p}$  &  $1.83 \pm 0.07~R_{\mathrm{Jup}}$ &1\\
    $T_{\rm eq}$  & 2260 $\pm$ 50~K &1\\
     \noalign{\smallskip}
\hline
 \noalign{\smallskip}
& Orbital parameters   & \\
 \noalign{\smallskip}
\hline
 \noalign{\smallskip}
     $T_0$  & $7909.5906^{+0.0003}_{-0.0002}$~BJD-2450000 &1\\
     $P$  & $3.474119^{+0.000005}_{-0.000006}$~days &1\\
     $e$ & 0 & fixed \\
     $K_{\rm s}$ & 322.51$^*$ \ms & 4\\  
    $K_{\rm p}$ & 175$^*$ \kms & \\ 
     $V_{\rm sys}$  & $-24.48 \pm 0.04$~\kms & 2\\
      \noalign{\smallskip}
\hline    
  \multicolumn{3}{l}{$^{*}$\footnotesize{Assuming the upper mass limit for the planet.}} \\                              
\end{tabular}
\tablebib{$^1$~\cite{Talens2018}; $^2$~\cite{rainer}; $^3$~\cite{Lund2017}; $^4$~\cite{2019A&A...628A...9C}}

\end{table}

\section{Statistics\label{sec:statistics}}

\begin{figure}[!ht]
\centering
\includegraphics[width=\linewidth]{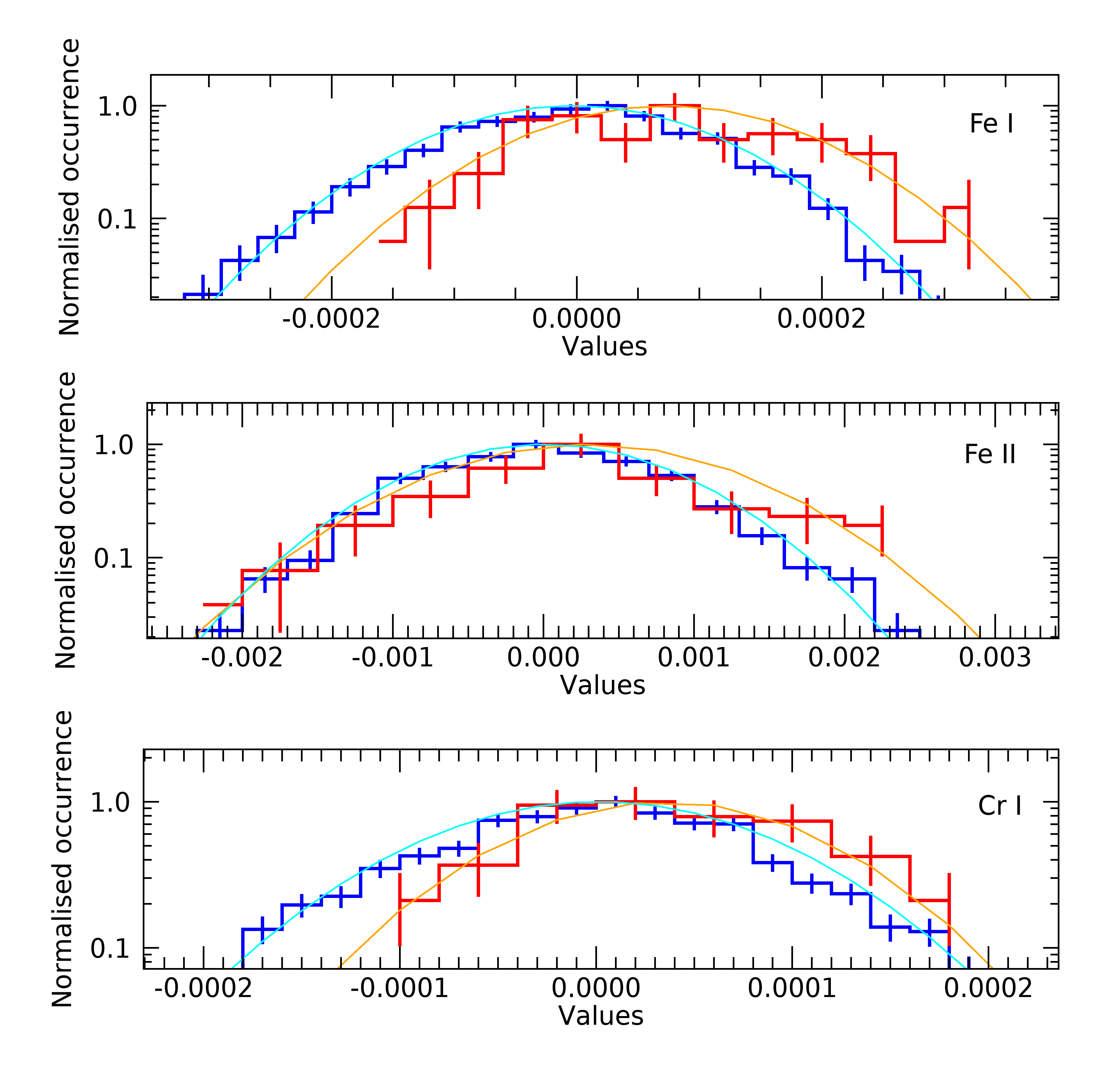}
\caption{Distribution of the night 1 in-trail and out-of-trail samples for \ion{Fe}{i}, \ion{Fe}{ii}, and \ion{Cr}{i}, respectively.
The blue histogram represents the out-of-trail distribution, while the red histogram the in-trail one. The respective Gaussian distributions are 
shown as cyan and orange lines.
}
\label{fig:signi_FeI}
\end{figure}

\begin{figure}[!ht]
\centering
\includegraphics[width=\linewidth]{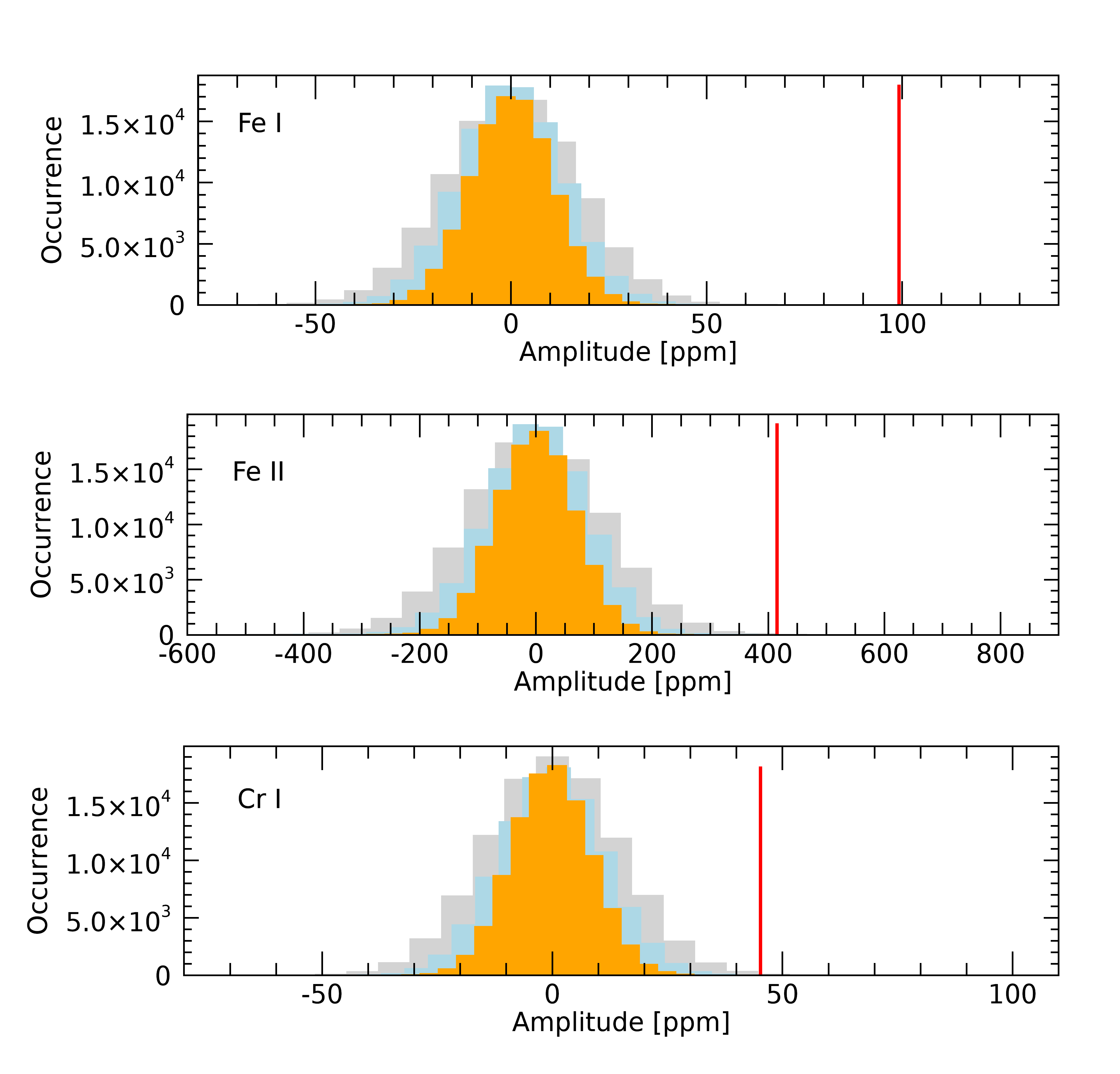}
\caption{Distributions created with the bootstrap method for \ion{Fe}{i}, \ion{Fe}{ii}, and \ion{Cr}{i}, respectively.
The gray, light blue and orange distributions are coming from fixed widths of 5, 10 and 20 \kms, respectively. 
The vertical red lines show the amplitude of the detection.
}
\label{fig:bootstrap}
\end{figure}

\begin{figure}[!ht]
\centering
\includegraphics[width=\linewidth]{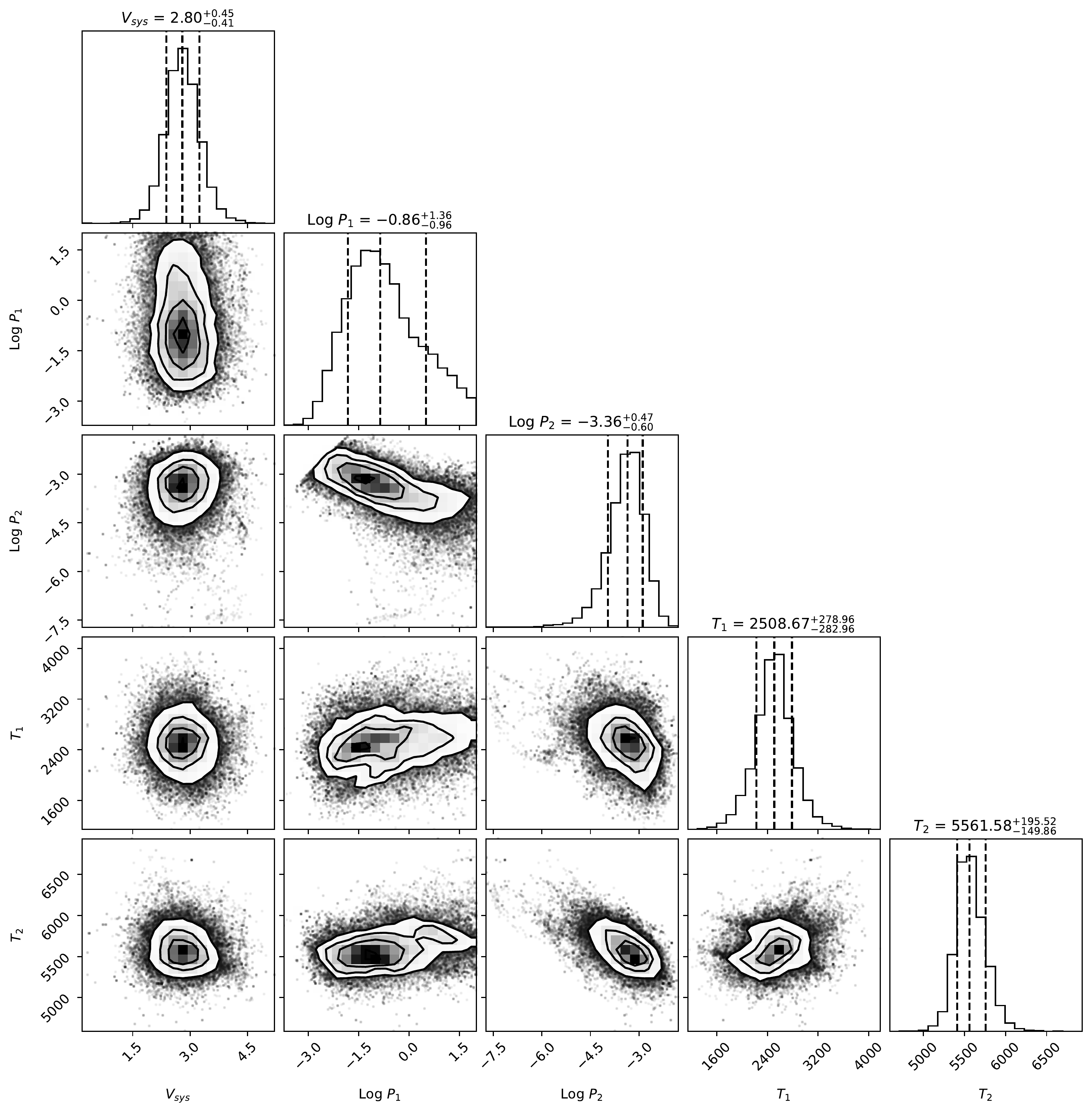}
\caption{Corner plot representing the posterior distribution of variables used for the DE-MCMC computations of the T-P profile parameters for night 1.
}
\label{fig:N1}
\end{figure}

\begin{figure}[!ht]
\centering
\includegraphics[width=\linewidth]{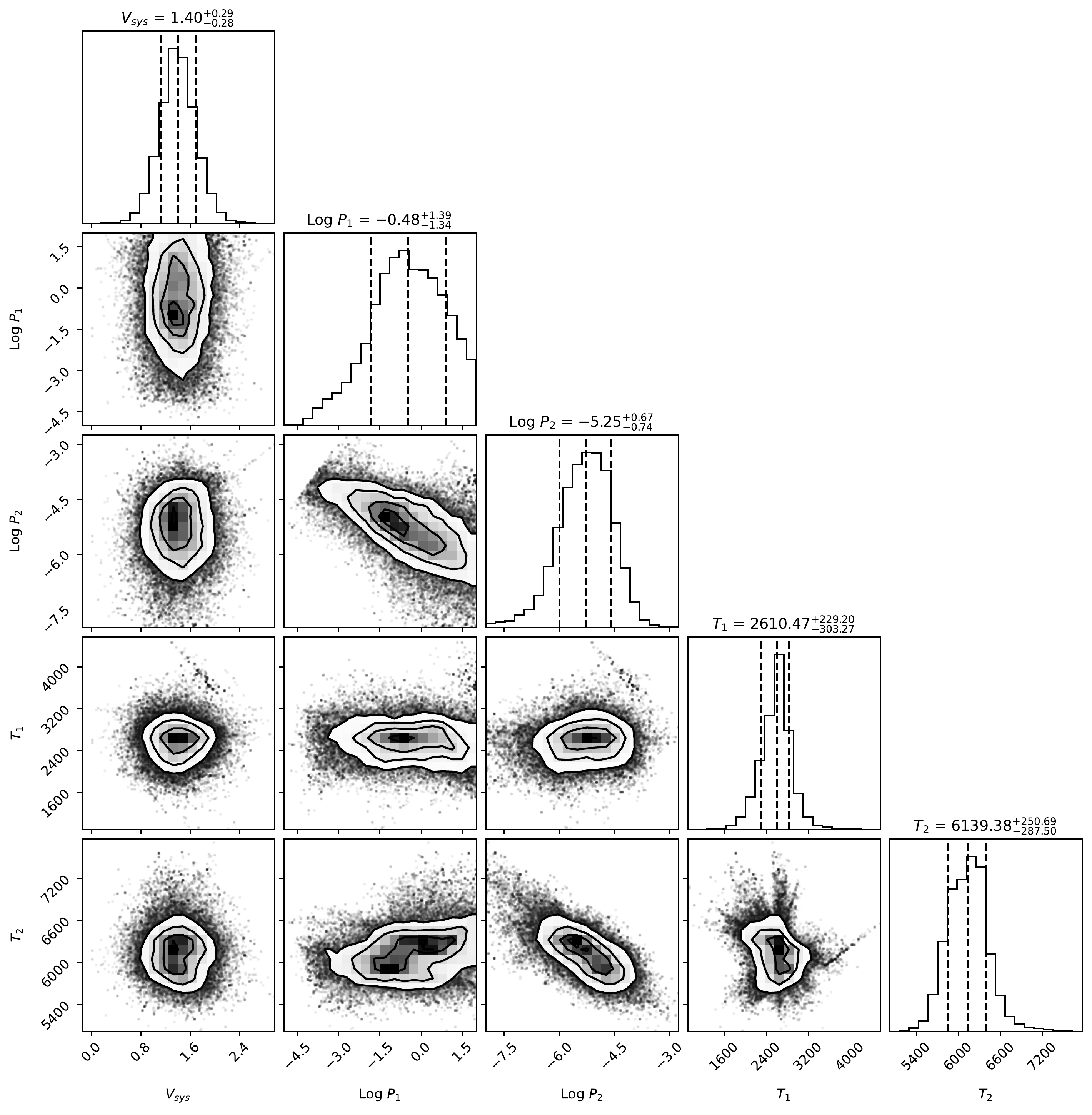}
\caption{Corner plot representing the posterior distribution of variables used for the DE-MCMC computations of the T-P profile parameters for night 2.
}
\label{fig:N2}
\end{figure}

\end{appendix}

\begin{thebibliography}{}

\bibitem[Arcangeli et al.(2018)]{2018ApJ...855L..30A} Arcangeli, J., D{\'e}sert, J.-M., Line, M.~R., et al.\ 2018, \apjl, 855, L30
\bibitem[Arcangeli et al.(2021)]{2021A&A...646A..94A} Arcangeli, J., D{\'e}sert, J.-M., Parmentier, V., et al.\ 2021, \aap, 646, A94.
\bibitem[Baxter et al.(2020)]{2020A&A...639A..36B} Baxter C., D{\'e}sert J.-M., Parmentier V., et al., 2020, A\&A, 639, A36.
\bibitem[Bell \& Cowan(2018)]{2018ApJ...857L..20B} Bell, T.~J. \& Cowan, N.~B.\ 2018, \apjl, 857, L20
\bibitem[Birkby et al.(2017)]{2017AJ....153..138B} Birkby, J.~L., de Kok, R.~J., Brogi, M., et al.\ 2017, \aj, 153, 138
\bibitem[Borsa et al.(2015)]{borsa} Borsa, F., Scandariato, G., Rainer, M., et al.\ 2015, \aap, 578, A64 
\bibitem[Borsa et al.(2019)]{2019A&A...631A..34B} Borsa, F., Rainer, M., Bonomo, A.~S., et al.\ 2019, \aap, 631, A34
\bibitem[Borsa et al.(2021)]{2021A&A...645A..24B} Borsa, F., Allart, R., Casasayas-Barris, N., et al.\ 2021, \aap, 645, A24
\bibitem[Brogi et al.(2018)]{2018A&A...615A..16B} Brogi, M., Giacobbe, P., Guilluy, G., et al.\ 2018, \aap, 615, A16
\bibitem[Cabot et al.(2019)]{2019MNRAS.482.4422C} Cabot, S.~H.~C., Madhusudhan, N., Hawker, G.~A., et al.\ 2019, \mnras, 482, 4422.
\bibitem[Casasayas-Barris et al.(2018)]{2018A&A...616A.151C} Casasayas-Barris, N., Pall{\'e}, E., Yan, F., et al.\ 2018, \aap, 616, A151
\bibitem[Casasayas-Barris et al.(2019)]{2019A&A...628A...9C} Casasayas-Barris, N., Pall{\'e}, E., Yan, F., et al.\ 2019, \aap, 628, A9
\bibitem[Claudi et al.(2017)]{giarps} Claudi, R., Benatti, S., Carleo, I., et al.\ 2017, European Physical Journal Plus, 132, 364
\bibitem[Cont et al.(2022)]{2022A&A...657L...2C} Cont, D., Yan, F., Reiners, A., et al.\ 2022, \aap, 657, L2. 
\bibitem[Cosentino et al.(2014)]{2014SPIE.9147E..8CC} Cosentino, R., Lovis, C., Pepe, F., et al.\ 2014, \procspie, 9147, 91478C
\bibitem[Deming \& Knutson(2020)]{2020NatAs...4..453D} Deming, D. \& Knutson, H.~A.\ 2020, Nature Astronomy, 4, 453
\bibitem[Eastman et al.(2013)]{Eastmanetal2013} Eastman, J., Gaudi, B. S. \& Agol, E. 2013, \pasp, 125, 923
\bibitem[Ehrenreich et al.(2020)]{espresso1} Ehrenreich, D., Lovis, C., Allart, R., et al.\ 2020, \nat, 580, 597
\bibitem[Fortney et al.(2008)]{2008ApJ...678.1419F} Fortney, J.~J., Lodders, K., Marley, M.~S., et al.\ 2008, \apj, 678, 1419
\bibitem[Fossati et al.(2020)]{2020A&A...643A.131F} Fossati, L., Shulyak, D., Sreejith, A.~G., et al.\ 2020, \aap, 643, A131
\bibitem[Fossati et al.(2021)]{fossati2021} Fossati, L., Young, M.~E., Shulyak, D., et al.\ 2021, \aap, 653, A52
\bibitem[Fu et al.(2022)]{2022ApJ...925L...3F} Fu, G., Sing, D.~K., Lothringer, J.~D., et al.\ 2022, \apjl, 925, L3
\bibitem[Giacobbe et al.(2021)]{2021Natur.592..205G} Giacobbe, P., Brogi, M., Gandhi, S., et al.\ 2021, \nat, 592, 205
\bibitem[Gibson et al.(2020)]{gibson2020} Gibson, N.~P., Merritt, S., Nugroho, S.~K., et al.\ 2020, \mnras, 493, 2215
\bibitem[Guilluy et al.(2019)]{2019A&A...625A.107G} Guilluy, G., Sozzetti, A., Brogi, M., et al.\ 2019, \aap, 625, A107
\bibitem[Guilluy et al.(2020)]{2020A&A...639A..49G} Guilluy, G., Andretta, V., Borsa, F., et al.\ 2020, \aap, 639, A49
\bibitem[Hoeijmakers et al.(2019)]{2019arXiv190502096H} Hoeijmakers, H.~J., Ehrenreich, D., Kitzmann, D., et al.\ 2019, \aap, 627, A165 
\bibitem[Hoeijmakers et al.(2020)]{2020A&A...641A.120H} Hoeijmakers, H.~J., Cabot, S.~H.~C., Zhao, L., et al.\ 2020, \aap, 641, A120
\bibitem[Hoeijmakers et al.(2020)]{2020A&A...641A.123H} Hoeijmakers, H.~J., Seidel, J.~V., Pino, L., et al.\ 2020, \aap, 641, A123.
\bibitem[Hubeny et al.(2003)]{2003ApJ...594.1011H} Hubeny, I., Burrows, A., \& Sudarsky, D.\ 2003, \apj, 594, 1011
\bibitem[Kasper et al.(2021)]{kasper2021} Kasper, D.~H., Bean, J.~L., Line, M.~R., et al.\ 2021, arXiv:2108.08389
\bibitem[Kesseli et al.(2020)]{2020AJ....160..228K} Kesseli, A.~Y., Snellen, I.~A.~G., Alonso-Floriano, F.~J., et al.\ 2020, \aj, 160, 228
\bibitem[Lothringer et al.(2018)]{2018ApJ...866...27L} Lothringer, J.~D., Barman, T., \& Koskinen, T.\ 2018, \apj, 866, 27
\bibitem[Lothringer \& Barman(2019)]{2019ApJ...876...69L} Lothringer, J.~D. \& Barman, T.\ 2019, \apj, 876, 69
\bibitem[Lund et al.(2017)]{Lund2017} Lund, M.~B., Rodriguez, J.~E., Zhou, G., et al.\ 2017, \aj, 154, 194
\bibitem[Molli{\`e}re et al.(2019)]{pRT} Molli{\`e}re, P., Wardenier, J.~P., van Boekel, R., et al.\ 2019, \aap, 627, A67
\bibitem[Nugroho et al.(2017)]{2017AJ....154..221N} Nugroho, S.~K., Kawahara, H., Masuda, K., et al.\ 2017, \aj, 154, 221
\bibitem[Nugroho et al.(2020)]{2020ApJ...898L..31N} Nugroho, S.~K., Gibson, N.~P., de Mooij, E.~J.~W., et al.\ 2020, \apjl, 898, L31
\bibitem[Nugroho et al.(2020)]{Nugroho2020} Nugroho, S.~K., Gibson, N.~P., de Mooij, E.~J.~W., et al.\ 2020, \mnras, 496, 504
\bibitem[Parmentier et al.(2018)]{2018A&A...617A.110P} Parmentier, V., Line, M.~R., Bean, J.~L., et al.\ 2018, \aap, 617, A110
\bibitem[Pino et al.(2020)]{2020ApJ...894L..27P} Pino, L., D{\'e}sert, J.-M., Brogi, M., et al.\ 2020, \apjl, 894, L27
\bibitem[Rainer et al.(2021)]{rainer} Rainer, M., Borsa, F., Pino, L., et al.\ 2021, A\&A, A\&A, 649, A29
\bibitem[Scandariato et al.(2020)]{2020arXiv201210435S} Scandariato, G., Borsa, F., Sicilia, D., et al.\ 2021, A\&A, 646, A159
\bibitem[Snellen et al.(2008)]{2008A&A...487..357S} Snellen, I.~A.~G., Albrecht, S., de Mooij, E.~J.~W., \& Le Poole, R.~S.\ 2008, \aap, 487, 357 
\bibitem[Stangret et al.(2020)]{2020A&A...638A..26S} Stangret, M., Casasayas-Barris, N., Pall{\'e}, E., et al.\ 2020, \aap, 638, A26
\bibitem[Talens et al.(2018)]{Talens2018} Talens, G.~J.~J., Justesen, A.~B., Albrecht, S., et al.\ 2018, \aap, 612, A57
\bibitem[Ter Braak(2006)]{TerBraak2006} Ter Braak, C. J. F. 2006, Statistics and Computing, 16, 239
\bibitem[Vidal-Madjar et al.(2010)]{2010A&A...523A..57V} Vidal-Madjar, A., Arnold, L., Ehrenreich, D., et al.\ 2010, \aap, 523, A57 
\bibitem[Yan et al.(2020)]{2020A&A...640L...5Y} Yan, F., Pall{\'e}, E., Reiners, A., et al.\ 2020, \aap, 640, L5
\bibitem[Yan et al.(2022)]{2022arXiv220108759Y} Yan, F., Reiners, A., Pall{\'e}, E., et al.\ 2022, arXiv:2201.08759

\end{thebibliography}
\end{document}